\documentclass{WileyMSP-template}

\usepackage{graphicx}
\usepackage{dcolumn}
\usepackage{bm}

\usepackage[utf8]{inputenc}
\usepackage[T1]{fontenc}
\usepackage{mathptmx}
\usepackage{etoolbox}

\usepackage{textcomp, gensymb}
\usepackage{amssymb}
\usepackage{verbatim}
\usepackage{subcaption}
\usepackage{float}
\usepackage{booktabs}
\usepackage{wrapfig}
\usepackage{xcolor}
\usepackage{mathtools}
\usepackage{makecell}
\usepackage{tabularx}
\usepackage{multirow}
\usepackage{multicol}
\usepackage{siunitx}
\newcommand{\SIadj}[2]{\SI[number-unit-product={\text{-}}]{#1}{#2}}

\makeatletter
\newcommand{\RB}{\@ifstar{\@Rb}{\@Ra}}
\newcommand{\@Rb}[1]{\textcolor[HTML]{0da34e}{[#1]}} 
\newcommand{\@Ra}[1]{\textcolor[HTML]{0da34e}{[\textbf{RB:} #1]}}
\makeatother

\begin{document}

\pagestyle{fancy}
\rhead{\includegraphics[width=2.5cm]{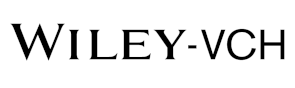}}


\title{Tunable Epsilon Near Zero Metamaterial with Rotating Obround-Shaped Meta-Atoms}

\maketitle


\author{Rustam~Balafendiev$^{\dagger}$},
\author{Gagandeep Kaur$^{\dagger}$},
\author{Jim A. Enriquez},
\author{Gaganpreet Singh},
\author{Alexander~J.~Millar},
\author{Jon E. Gudmundsson},
\author{Pavel~Belov}



\begin{affiliations}
R. Balafendiev, J. E. Gudmundsson\\
Science Institute, University of Iceland, Dunhagi 5, 107 Reykjavik, Iceland\\
Email Address: rub8-at-hi-dot-is

G. Kaur, G. Singh, J. E. Gudmundsson\\
The Oskar Klein Centre, Department of Physics, Stockholm University, AlbaNova, SE-10691 Stockholm, Sweden\\

R. Balafendiev, J. A. Enriquez, P. Belov\\
ITMO University, 197101 St. Petersburg, Russia

A. J. Millar\\
Theoretical Physics Division, Fermi National Accelerator Laboratory, Batavia, IL 60510, USA\\
Superconducting Quantum Materials and Systems Center (SQMS), Fermi National Accelerator Laboratory, Batavia, IL 60510, USA

\end{affiliations}

\medskip
\noindent $^{\dagger}$ These authors contributed equally.
   

\begin{abstract}


A new design of a microwave-range ENZ metamaterial consisting of rods with an obround cross-section is proposed. The plasma frequency of the metamaterial can be tuned by rotating the constituent meta-atoms. Tunability of the plasma frequency by $26 \%$ is demonstrated both experimentally and numerically. 
The observed tuning range is dramatically higher than in the one observed in natural materials at optical range.

\end{abstract}

\section{Introduction}

Epsilon-Near-Zero (ENZ) materials exhibit extremely low dielectric permittivity at specific frequencies, resulting in an infinite wavelength and typically a negligible group velocity. This unique property gives rise to distinctive optical phenomena such as slow-light effects, enhanced field confinement, and strong electromagnetic wave-matter interactions \cite{kinsey2019near}. Due to these characteristics, ENZ materials have garnered significant attention for enabling diverse electromagnetic applications, including wave tunneling \cite{silveirinha_tunneling_2006, powell_nonlinear_2009, adams_funneling_2011}, high-directional emission \cite{enoch_metamaterial_2002, alu_epsilon-near-zero_2007, kim_role_2016}, and enhanced nonlinear effects \cite{fomra_nonlinear_2024}. Metals can naturally exhibit ENZ behavior near their plasma frequency in the UV-visible range \cite{Naik2013}, but high losses in this regime restrict their practical use. In contrast, transparent conductive oxides like Indium Tin Oxide (ITO) \cite{ni_property_2020} provide a more viable platform for mid-infrared ENZ applications due to their reduced losses \cite{Wang2019}.\\

Structured media, such as metamaterials allow access to new spectral regions and offer enhanced degrees of freedom for achieving ENZ materials. By designing metal-dielectric structures—such as waveguides or wire media, ENZ material behavior can be effectively emulated \cite{rotman_plasma_1962}. Low-loss ENZ responses have been demonstrated across a wide spectral range, from the visible to the near-infrared, through configurations based on metal-dielectric layers \cite{suresh_enhanced_2021, genchi_tunable_2021}. Additionally, metallic waveguides operating near their cutoff frequencies exhibit ENZ behavior in both the visible and microwave regimes, enabling applications like enhanced radiative optical density of states \cite{vesseur_experimental_2013} and energy squeezing \cite{edwards_experimental_2008}.\\

To enable practical applications across various spectral ranges, it is essential to tune the frequency at which the ENZ condition occurs. Metal transition nitrides offer significant tunability of plasma frequency but also present high losses, which restricts their applicability \cite{bagheri_large-area_2015, popovic_low-loss_2017, patsalas_conductive_2018, patsalas_zirconium_2019}. Other tunable ENZ materials have been developed, including electrical tuning to alter the carrier concentration in ITO \cite{park_electrically_2015, ma_electrically_2024}, modifying the form factor of ITO nanowires \cite{zhou_broadband_2018}, and employing optical fields in a self-assembled liquid crystal–nanoparticle system \cite{turpin_reconfigurable_2014}. Metamaterials present a promising alternative, as they can be designed to achieve the ENZ condition and tailored for tunability \cite{turpin_reconfigurable_2014}. Additionally, metamaterials enable ENZ behavior in spectral regions, such as the microwave regime, where natural materials are unfeasible. This capability makes metamaterials particularly suited to meet the demands of practical applications.

Wire media (WM) have been extensively studied for their unique properties as metamaterials, particularly for their behavior as an artificial plasma and tunability \cite{rotman_plasma_1962, Brown, pendry_low_1998, Belov}. Previous research has explored the use of WM to exploit the ENZ condition for applications such as beam collimation \cite{shen_metamaterial-based_2016}, broadband microstrip antennas \cite{jafargholi_broadband_2015}, and enhancements in the gain and bandwidth of microstrip monopole antennas \cite{abdelgwad_high_2019}. Furthermore, the plasma frequency of WM has been shown to be tunable through various methods, including temperature variation \cite{Gorlach16}, the incorporation of ferrites \cite{Yang12}, and mechanical adjustments \cite{Ivzhenko2016, Kowitt23}. Among these tuning methods, mechanical tuning is particularly noteworthy, as it remains viable under conditions where other mechanisms—such as cryogenic temperatures or high magnetic fields—may be impractical. This makes mechanical tuning especially relevant in novel applications, such as the recently proposed use of WM for constructing plasma haloscopes aimed at dark matter detection \cite{Lawson19, Millar23}.

Mechanically tunable plasma haloscopes have been demonstrated using a tuning rod \cite{Boutan18, Zhong18}, while alternative mechanisms, including movable cavity walls \cite{McAllister24, Golm2023, Braggio23}, multiple rods in coupled cavities \cite{Jeong20}, and a design that utilizes a tunable plasmonic crystal of sapphire rods \cite{Bae23}, have been proposed. In our earlier work \cite{Balafendiev2022}, we demonstrated that a WM-filled cavity, due to the extremely large wavelength achieved at the ENZ condition, can harness a volume comparable to that of a conventional cavity operating at much lower frequencies, while maintaining a sufficiently high quality factor. 
In another work we have shown that the plasma frequency of the WM can be mechanically tuned, achieving a tuning range of up to 16\% \cite{Kowitt23}. However, this method does not fully preserve volume and the linear motion it relies upon is challenging to implement in a microwave cavity.

In this paper, we introduce a novel approach to mechanically tune the plasma frequency of WM by rotating pairs of wires with elongated cross-sections in the microwave regime. The metamaterial properties are examined through theoretical analysis, simulations, and experimental validation. By studying the resonance frequency of the TM fundamental mode in a metallic cavity filled with WM, we determine the plasma frequency and evaluate its tunability. This method enables substantial mechanical tuning of the plasma frequency, facilitating dynamic control of the ENZ condition.

\section{Results}
\subsection{Numerical Simulations of an Infinite Medium}

\begin{figure}
\centering
\includegraphics[width=.95\linewidth]{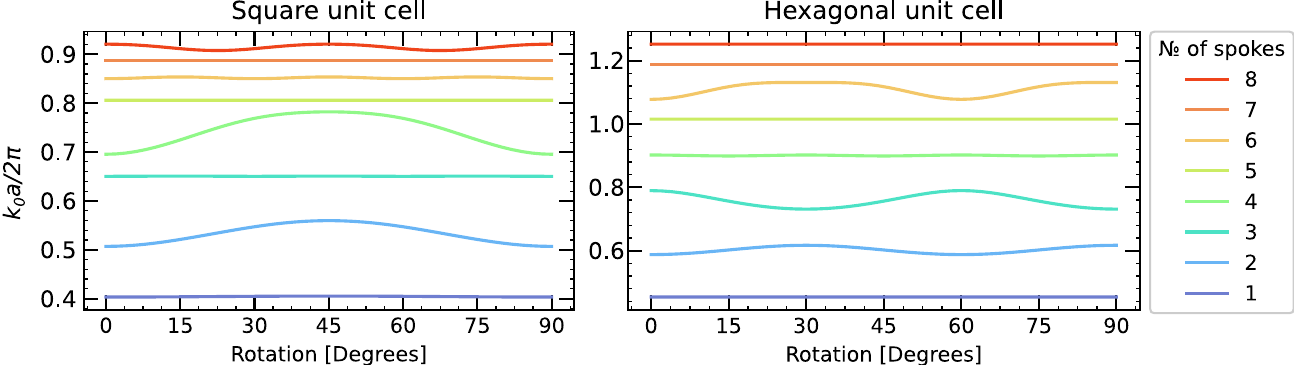}
\caption{Normalised plasma frequency as a function of the angle of rotation of the metaatom for two unit cells: square (left) and hexagonal(right). As can be seen, the cases with largest tuning share symmetry with the unit cell..
}
\label{fig:spokes}
\end{figure}

The plasma-like properties of a wire metamaterial are closely linked to the mutual inductance of its constituent wires. 
In a typical wire metamaterial, this inductance does not depend on the orientation of the wire within the unit cell as long as the wires remain parallel. However, for non-circular cross-sections, the inductance becomes linked to the orientation of individual elements.
Here, the periodic elements known as metaatoms are formed by the metallic spokes (see Table \ref{tab:spokes}). The size of the spokes is intentionally enlarged to exaggerate the effect. 
Depending on the symmetry of the metaatom relative to the symmetry of the unit cell, the tuning range varies between virtually zero and a little above ten percent. 
The tuning is largest when the symmetry of the metaatom matches that of the unit cell, such as the 2-, 4- and 8-spoked cases for the square cell and the 2-, 3- and 6-spoked cases for the hexagonal one (Figure \ref{fig:spokes}).

\begin{table}[h]
    \centering
    \begin{tabular}{cccccccccc}
    \hline
    \vspace{-3.5mm}\\
    & \makecell{Lowest mode electric\\field distribution}
    &\makecell{\includegraphics[width=0.06\columnwidth]{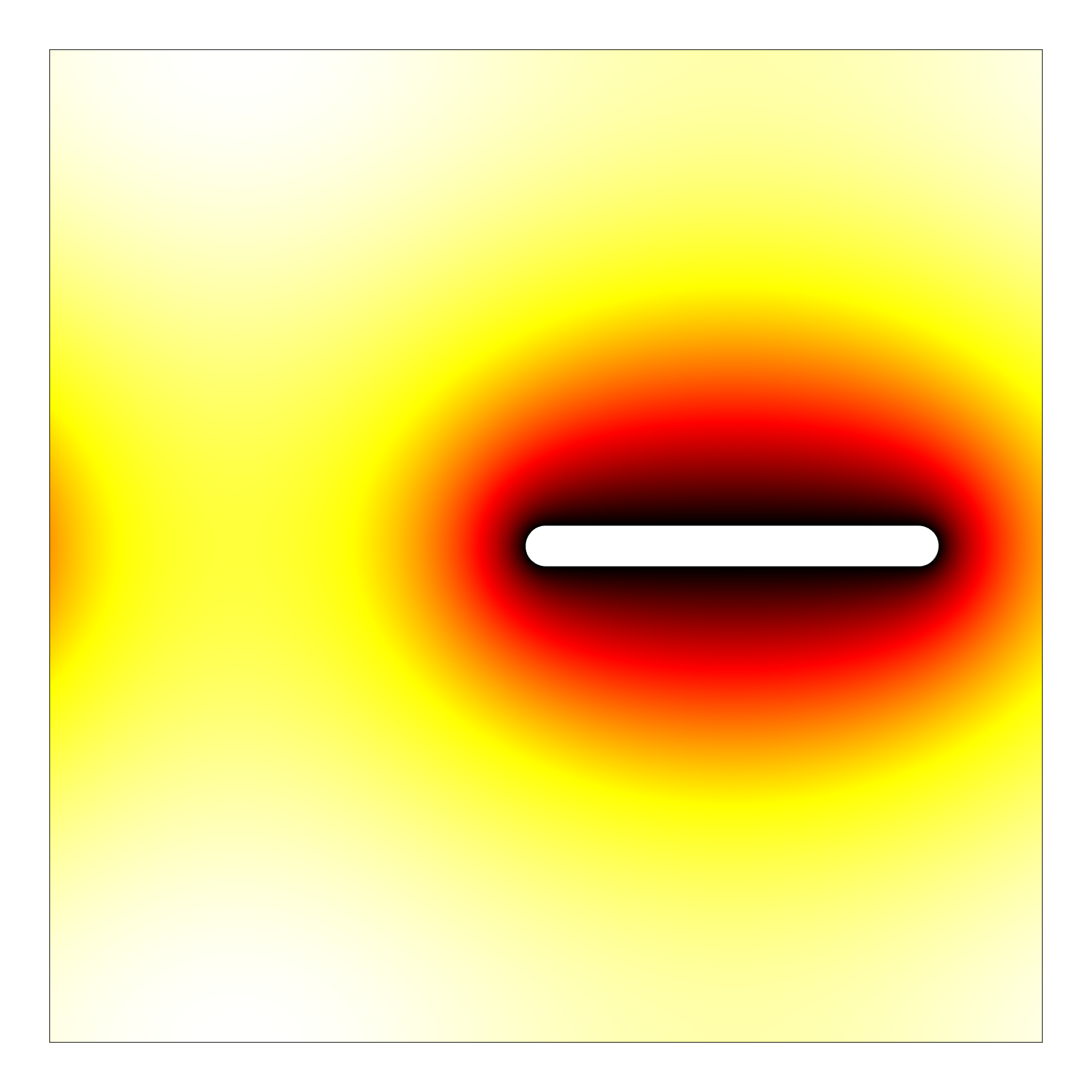}}
    &\makecell{\includegraphics[width=0.06\columnwidth]{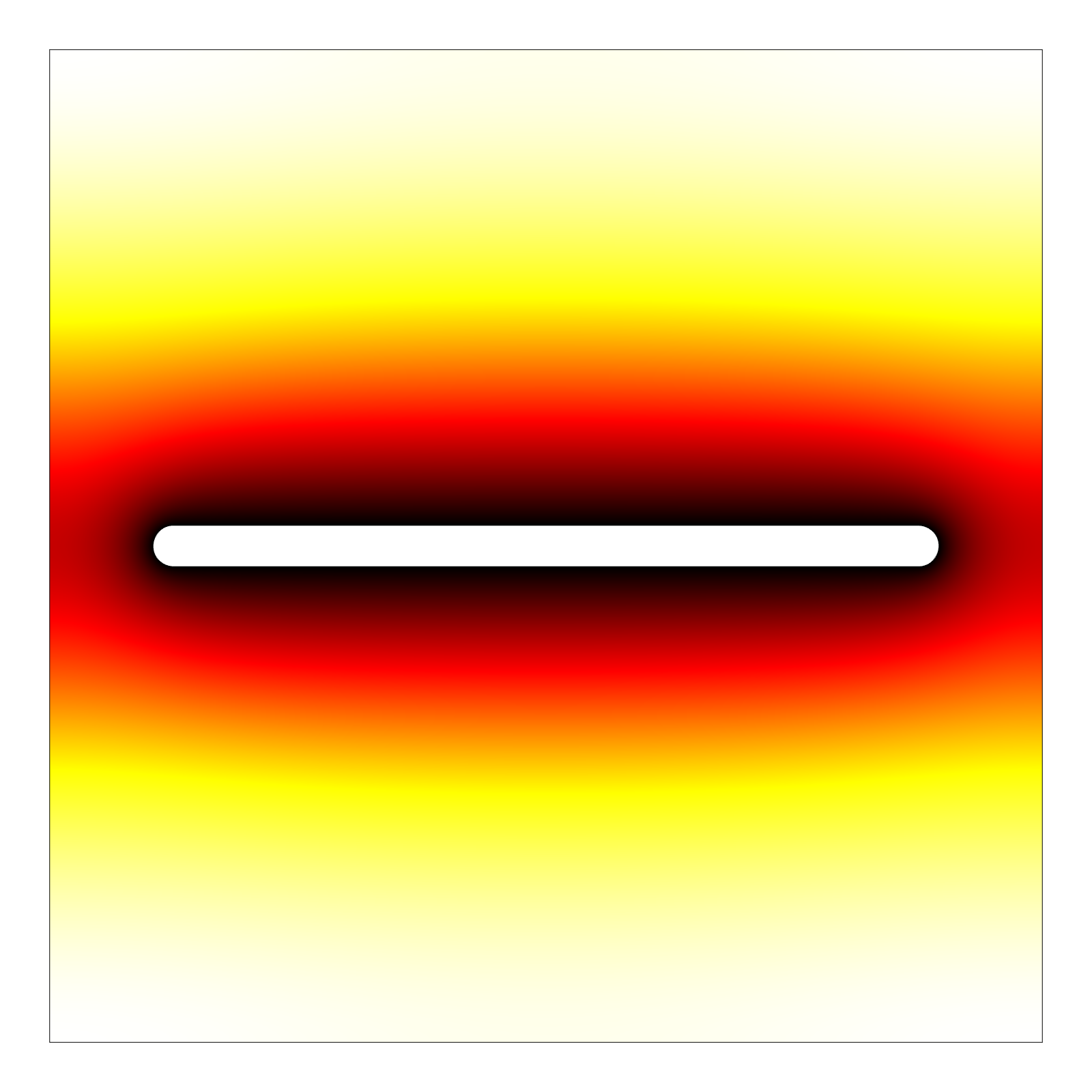}}
    &\makecell{\includegraphics[width=0.06\columnwidth]{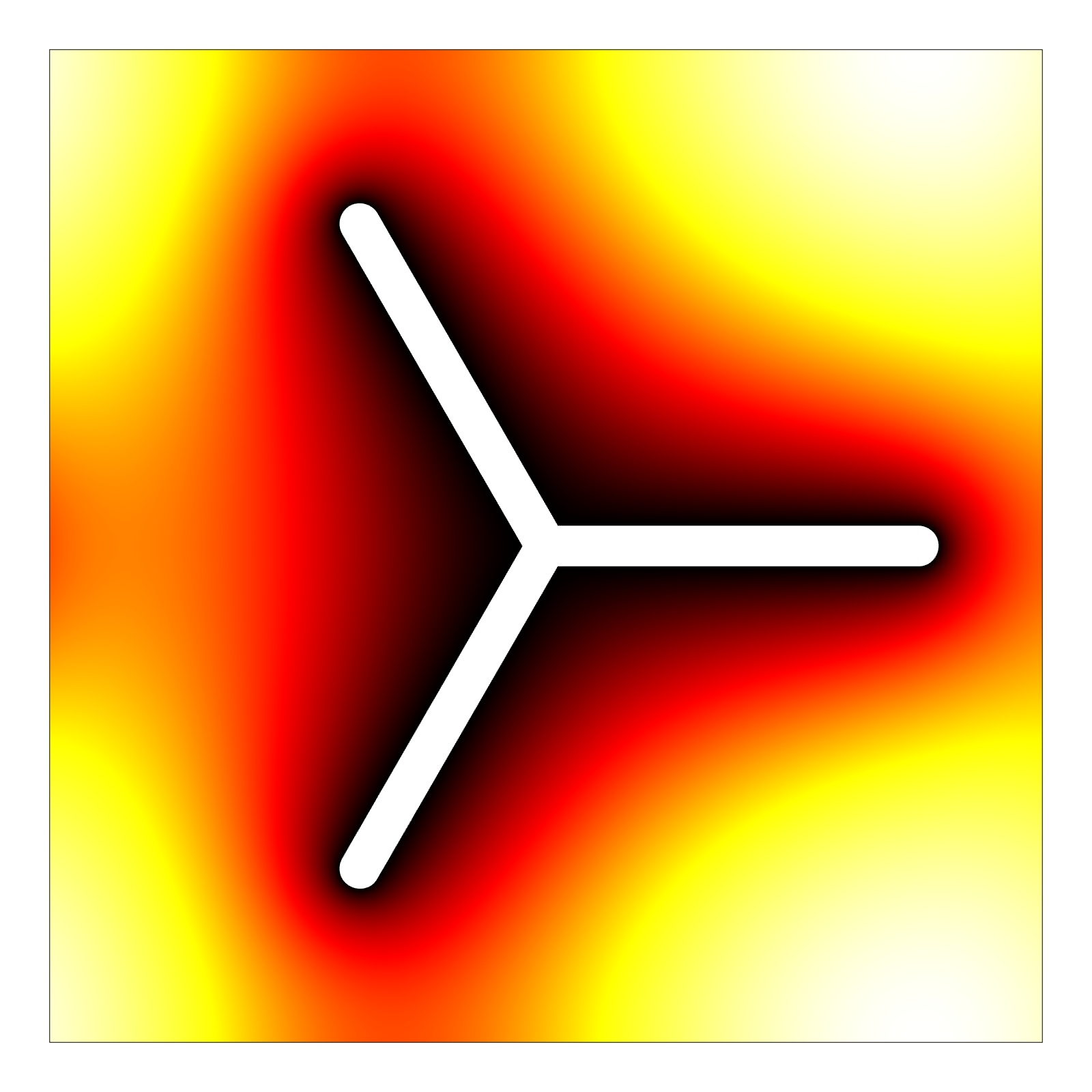}}
    &\makecell{\includegraphics[width=0.06\columnwidth]{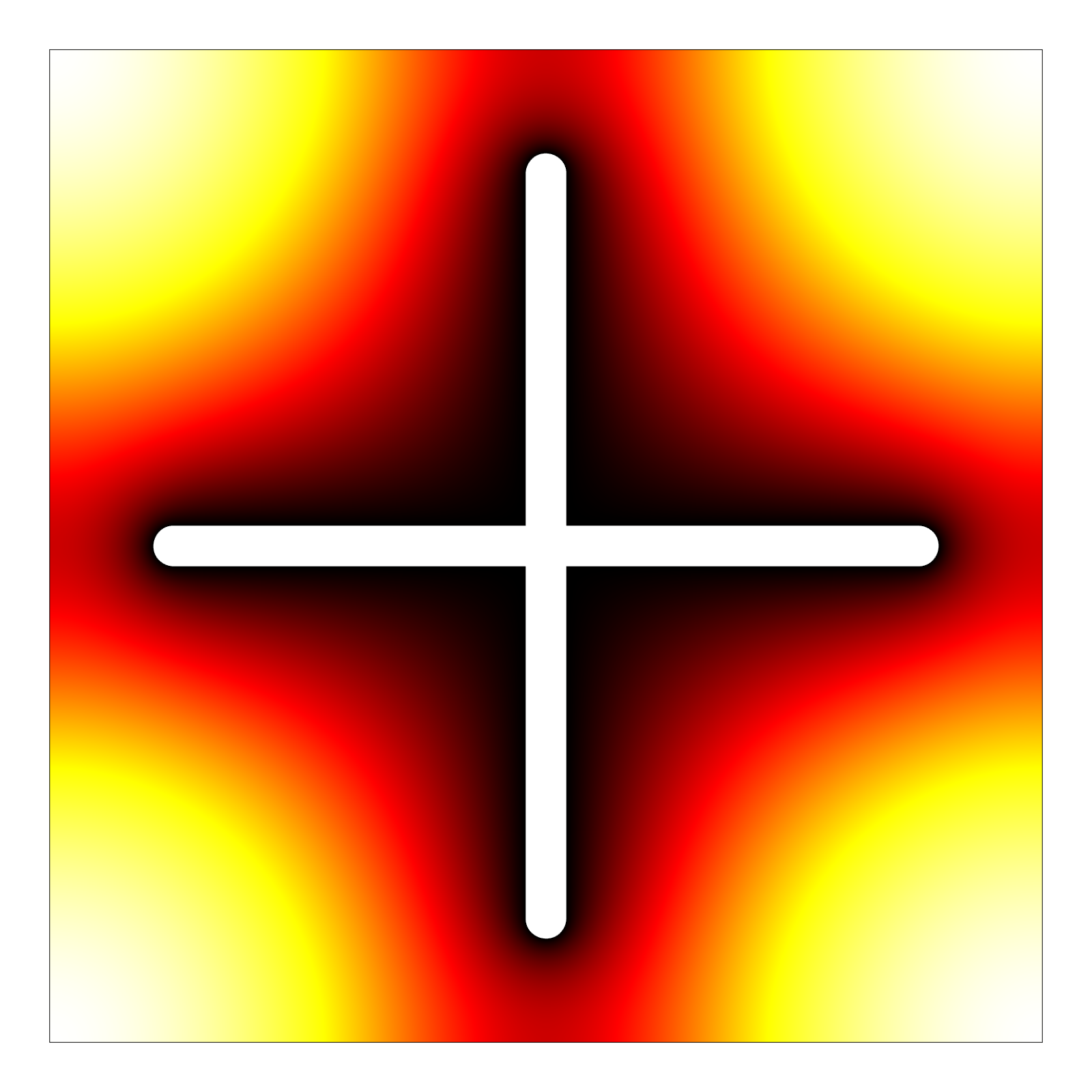}}
    &\makecell{\includegraphics[width=0.06\columnwidth]{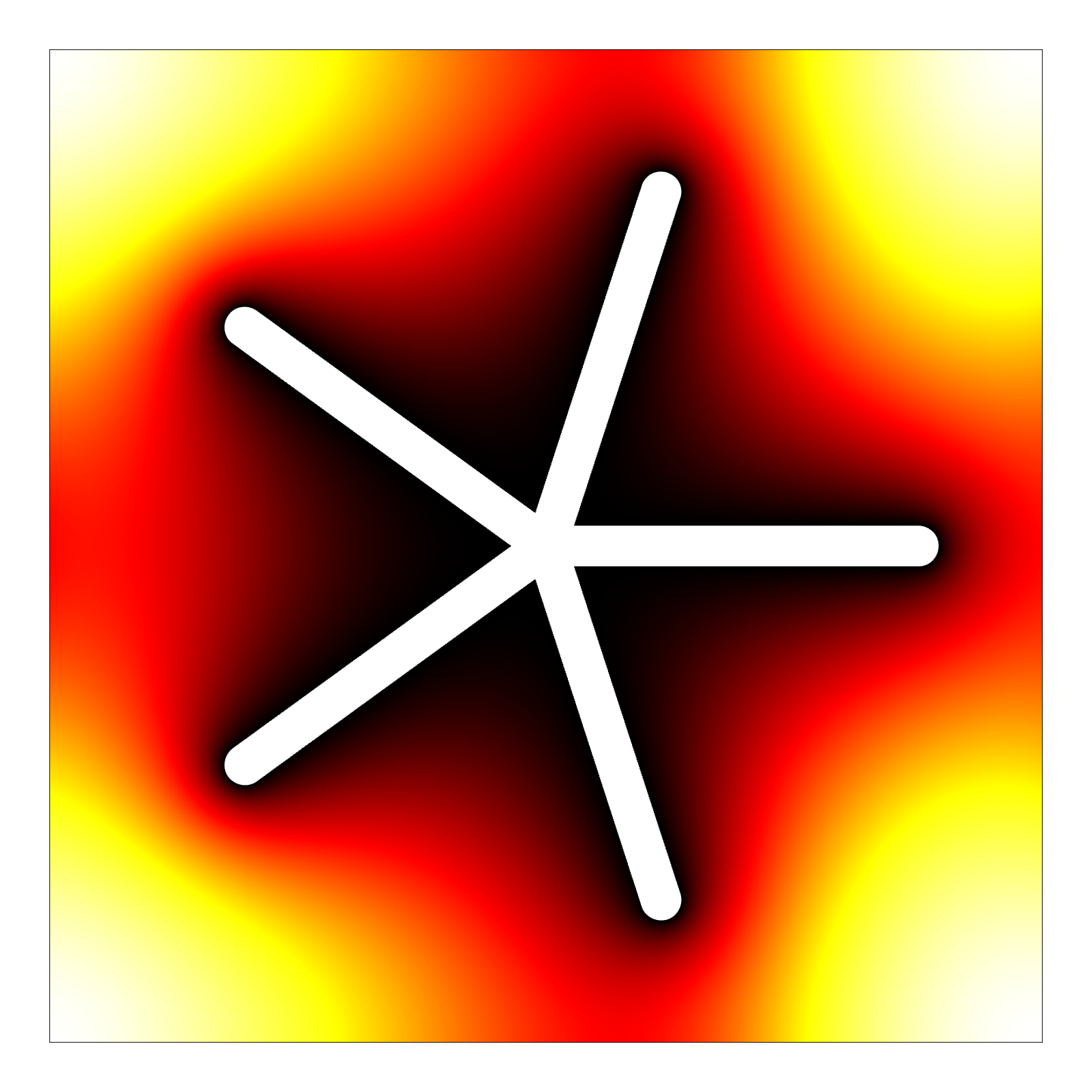}}
    &\makecell{\includegraphics[width=0.06\columnwidth]{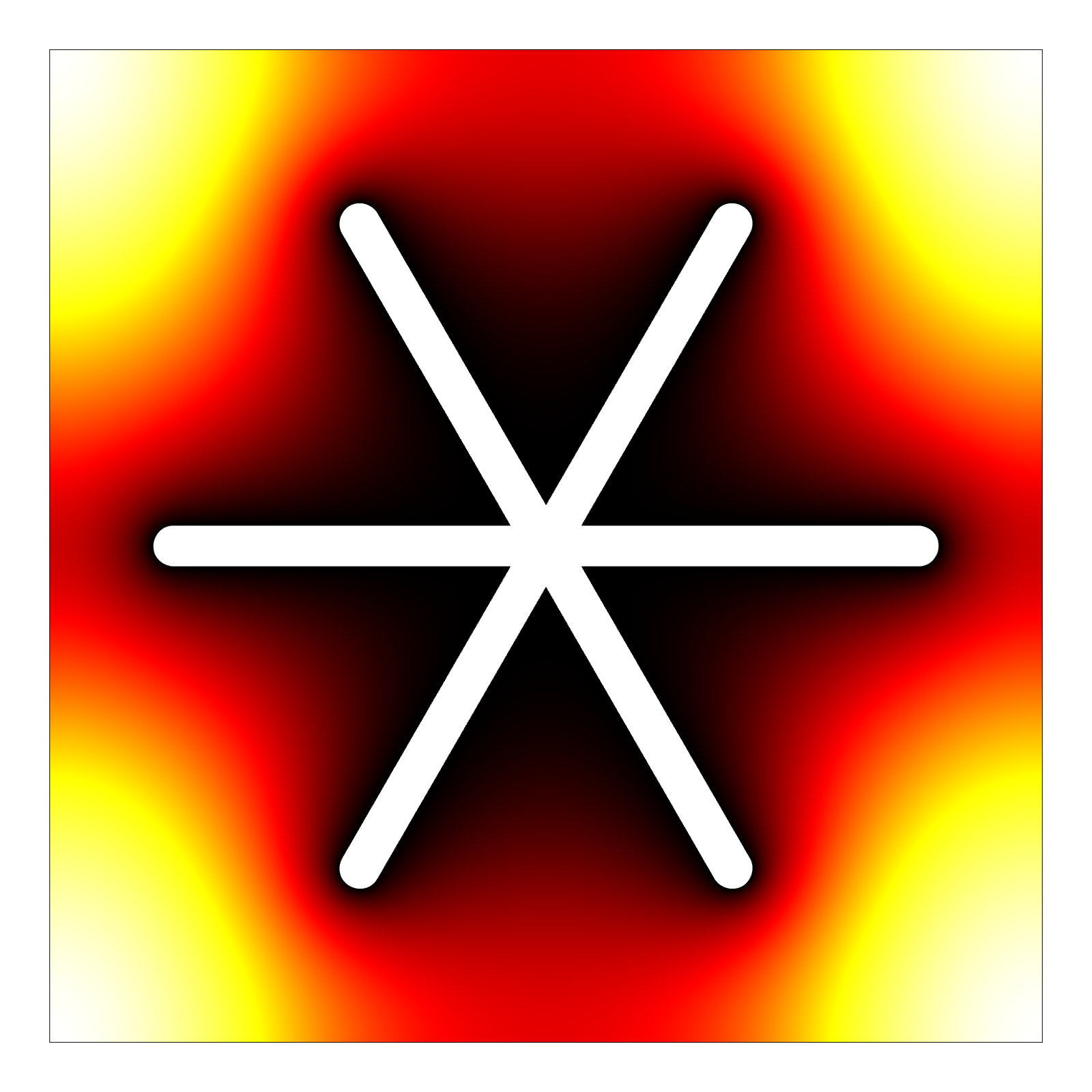}}
    &\makecell{\includegraphics[width=0.06\columnwidth]{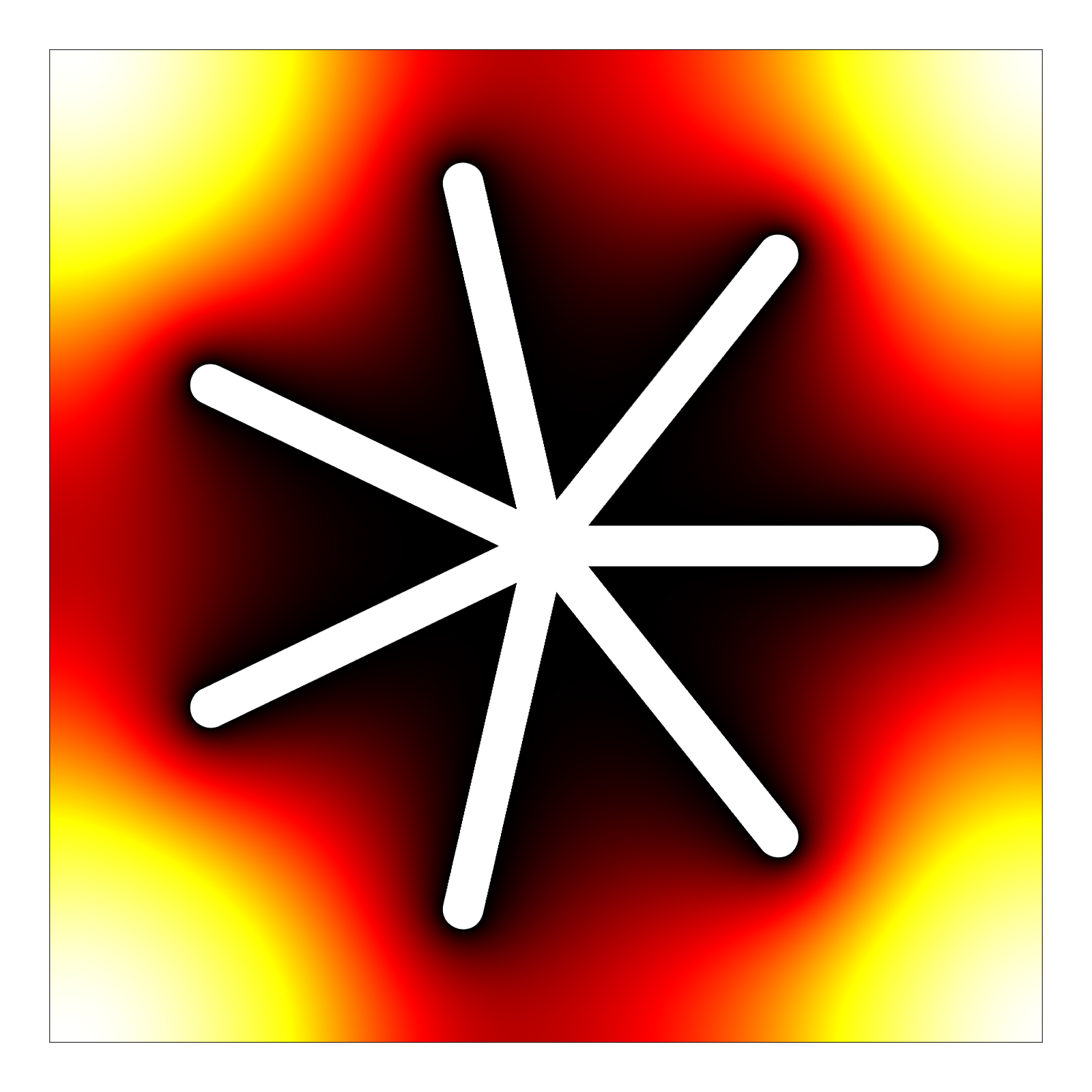}}
    &\makecell{\includegraphics[width=0.06\columnwidth]{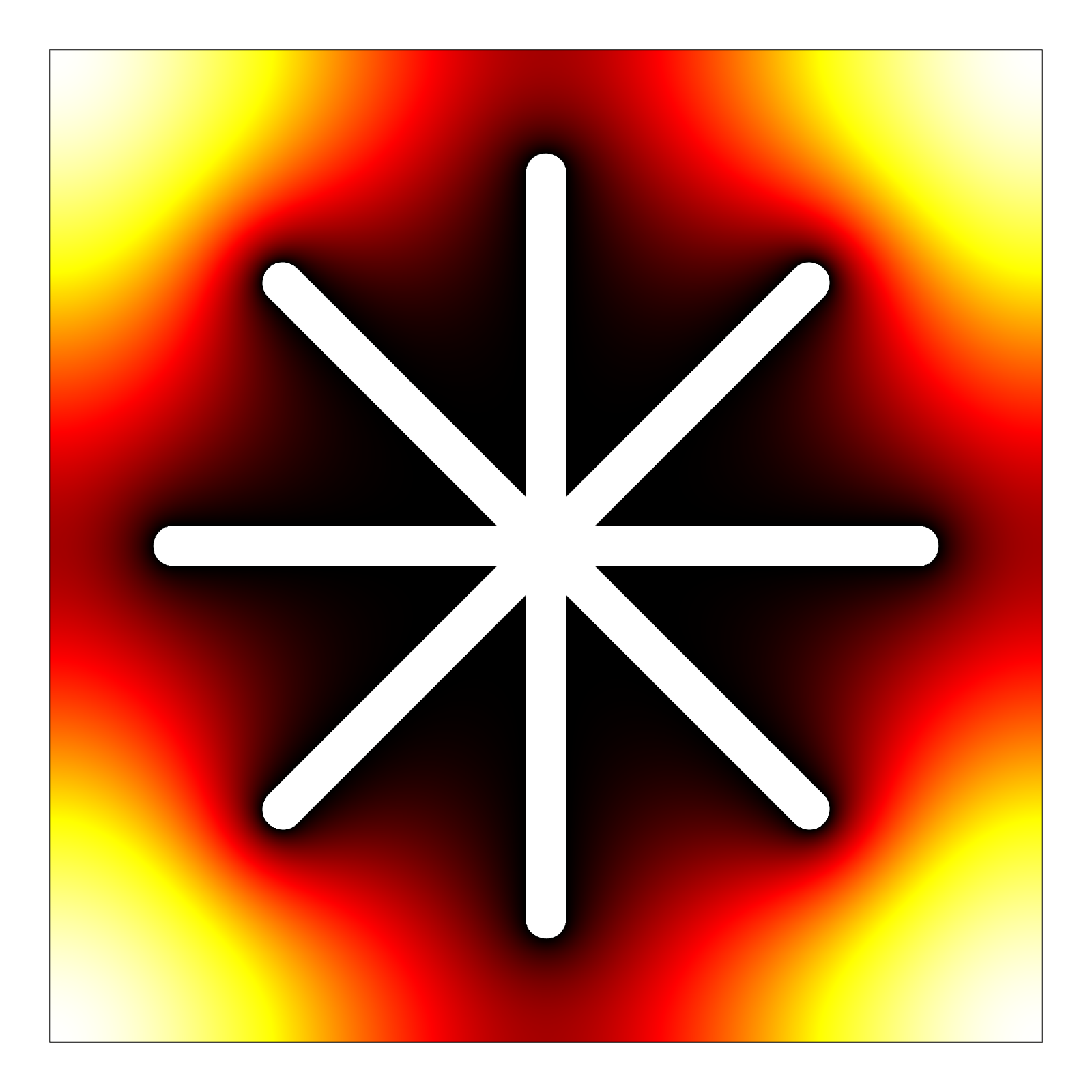}}\\
    \cline{2-10}
    & Lowest frequency, $k_0a/2\pi$ & 0.40 & 0.51 & 0.65 & 0.69 & 0.81 & 0.85 & 0.89 & 0.91 \\
    \multirow{-6}{*}{\rotatebox{90}{\parbox{2 cm}{\raggedright Square \\ unit cell}}}
    & Tuning range, \%  &  0.50 & 9.87 & 0.07 & 11.74  & 0.01  & 0.42  & 0.00 & 1.43 \\
    \hline
    \hline
    \vspace{-3.5mm}\\
    &\makecell{Lowest mode electric\\field distribution}
    &\makecell{\includegraphics[width=0.06\columnwidth]{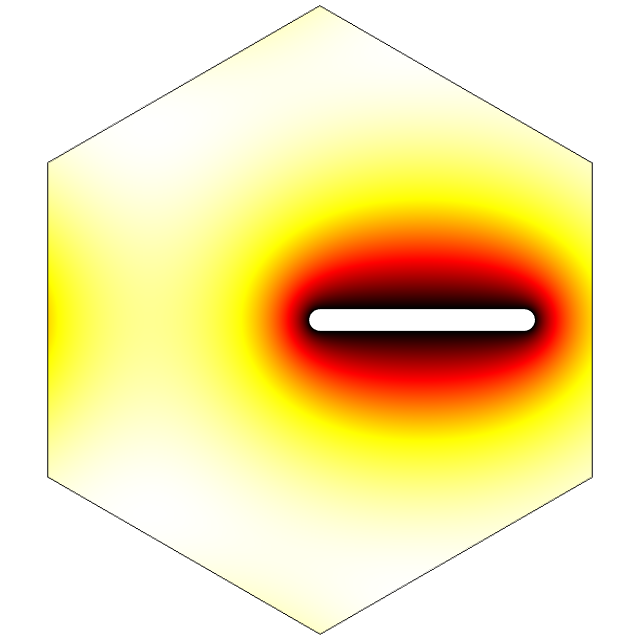}}
    &\makecell{\includegraphics[width=0.06\columnwidth]{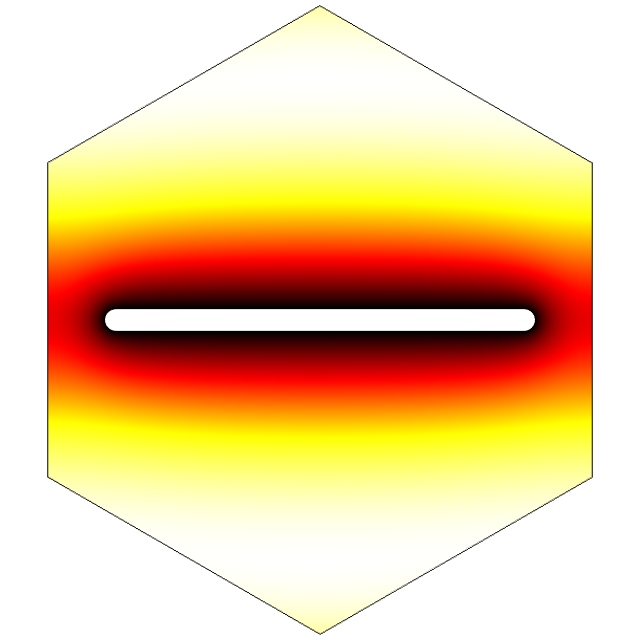}}
    &\makecell{\includegraphics[width=0.06\columnwidth]{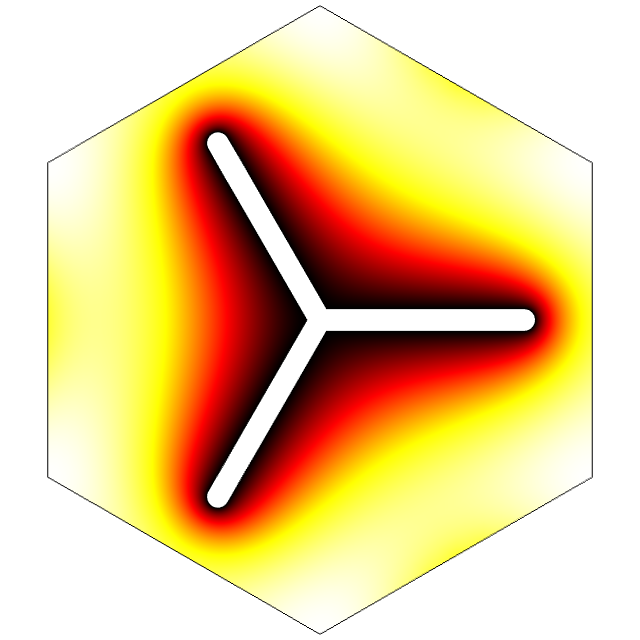}}
    &\makecell{\includegraphics[width=0.06\columnwidth]{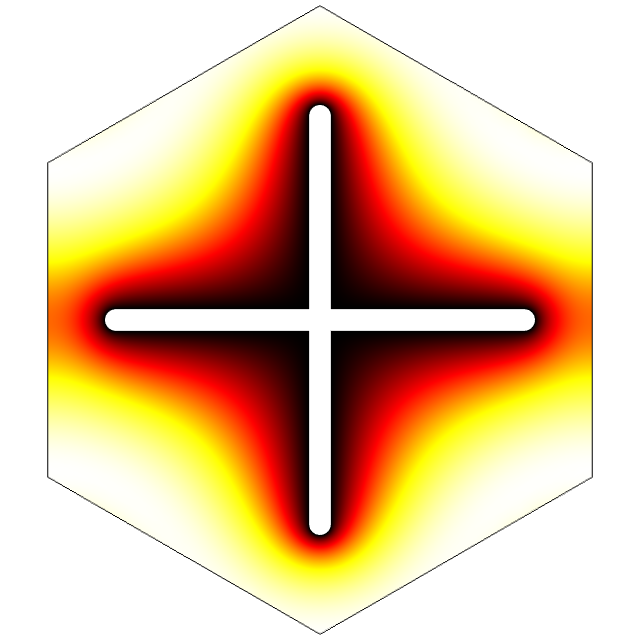}}
    &\makecell{\includegraphics[width=0.06\columnwidth]{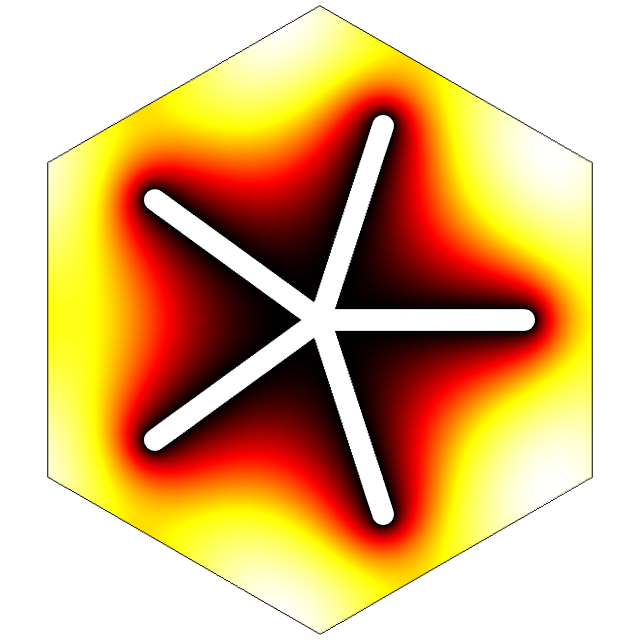}}
    &\makecell{\includegraphics[width=0.06\columnwidth]{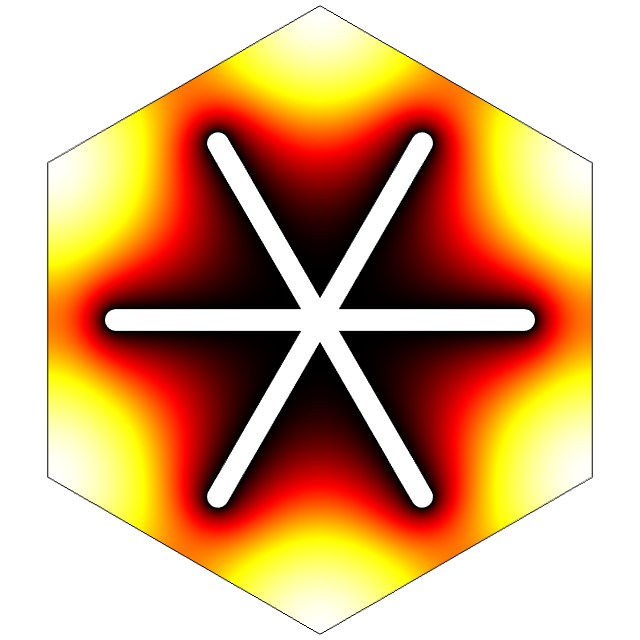}}
    &\makecell{\includegraphics[width=0.06\columnwidth]{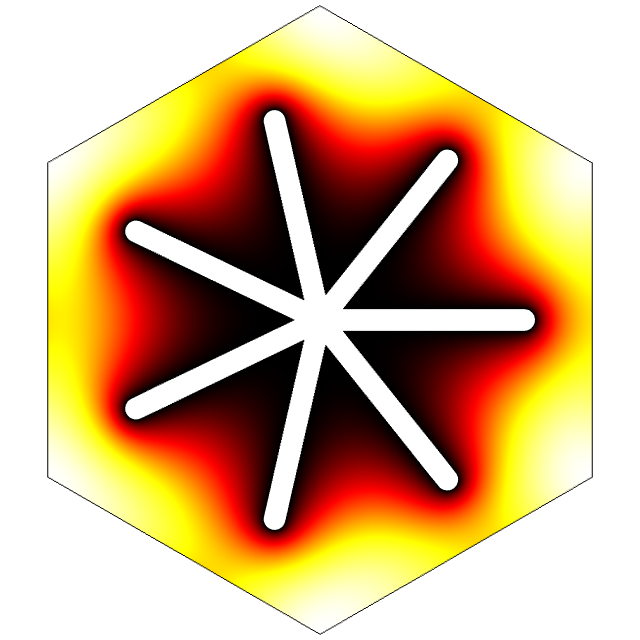}}
    &\makecell{\includegraphics[width=0.06\columnwidth]{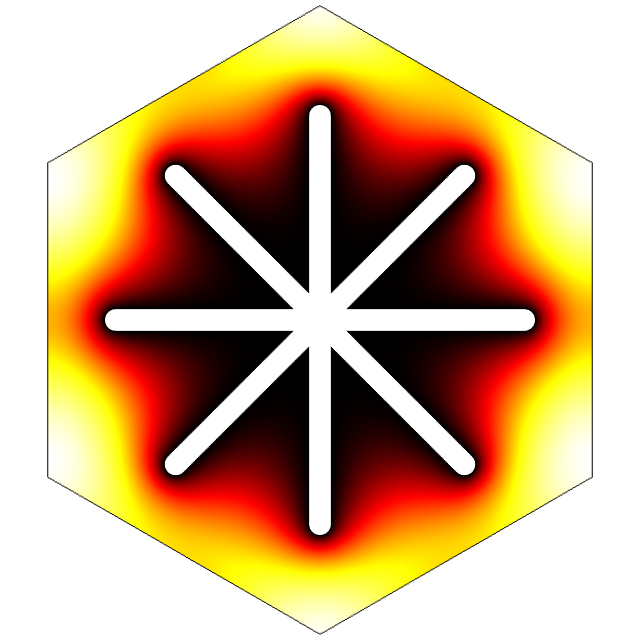}}\\
    \cline{2-10}
    &Lowest frequency, $k_0a/2\pi$ & 0.45 & 0.59 & 0.73 & 0.90 & 1.02 & 1.08 & 1.19 & 1.25 \\    \multirow{-6}{*}{\rotatebox{90}{\parbox{2 cm}{\raggedright Hexagonal \\ unit cell}}}
    &Tuning range, \%  &  0.08 & 4.91 & 7.72 & 0.33 & 0.01  & 4.85  & 0.00 & 0.01 \\
    \hline
    \end{tabular}
    \caption{Tuning characteristics for various spoke-based 2D unit cell geometries. The width of the unit cells is $a$, the diameter of the circle within which the spokes are placed is $3a/4$ and the width of the spokes themselves is $a/48$. The range of plasma frequency tuning is found by having the metaatom make a full rotation around the center of the unit center}
    \label{tab:spokes}
\end{table}

\begin{figure}[h]
\centering
\includegraphics[width=0.55\linewidth]{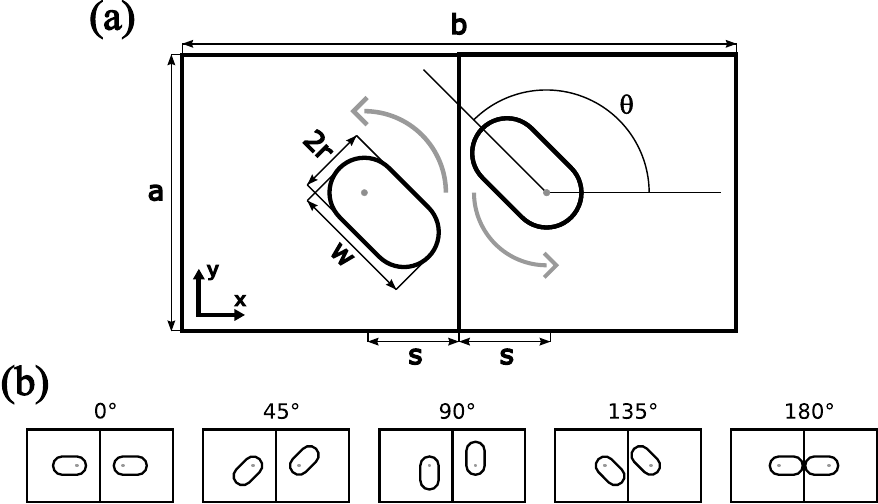}
\caption{\textbf{a)} A single unit cell of the proposed metamaterial with periods $a$ and $b$. The wires have a height of $2r$ and a width of $w$ and are rotated around the axis $s$ = $a/3$ away from the center by the angle $\theta$. \textbf{b)} Several configurations of the unit cell for different values of~$\theta$.}
\label{fig:uc}
\end{figure}

By intentionally breaking the symmetry of the unit cell and extending it to encompass two wires, the tunable range of the plasma frequency can be expanded. 
As shown in our previous work \cite{Kowitt23}, this range is maximized with the maximum possible change of distances between two neighboring elements, owing to the change of mutual inductance. 
The obround rod-based metamaterial is comprised of pairs of rotating eccentric metaatoms, capable of changing the mutual inductance by starting in a configuration not unlike that of a usual strip metamaterial and rotating closer to each other around the two neighboring sides (Figure \ref{fig:uc}a and c). In contrast to a regular wire medium, the unit cell of such a metamaterial is elongated, resulting in a rectangular cell with the width twice larger than the height. 

In this paper, we investigate the geometry of a specific case that consists of two squares with a side $a$, so that the width of the full unit cell is $b=2a$. Each half of the unit cell has a metallic obround rod with a width $w = 0.45 a$ and height of $2r~= 0.25a$, where $r$ is the radius of the semicircles comprising the sides of obround rod. The rods are rotated counterclockwise by the angle $\theta$ around the two axes placed at a distance $s=a/3$ away from the center of the unit cell.
At $\theta = 0^\circ$ the distance between the edges of the two rods equals $1.3a/6$ (Figure \ref{fig:uc}b). At $\theta = 180^\circ$ the rods are brought within $0.1a/6$ of each other. 

\begin{figure}
\includegraphics[width=1\linewidth]{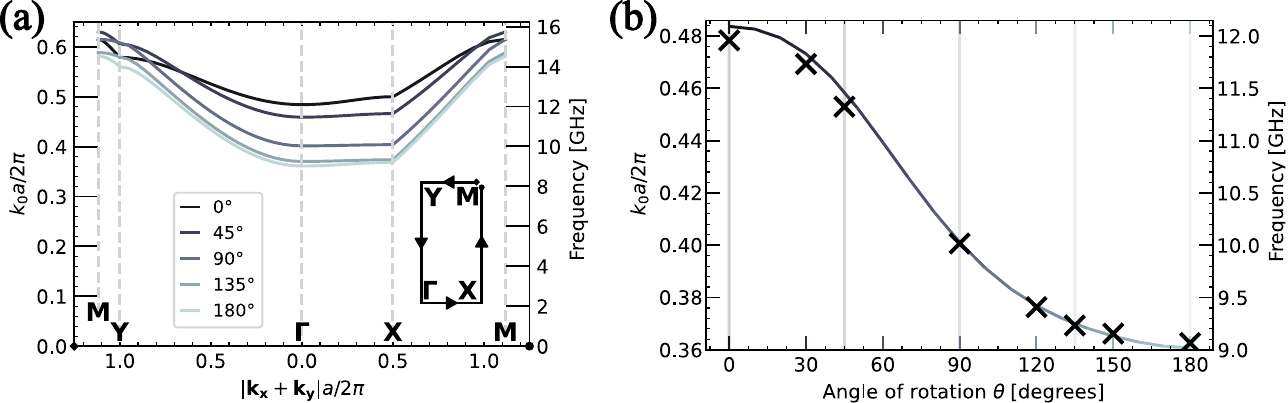}
\caption{\textbf{a)} Dispersion curves of the proposed metamaterial at several values of the rotation angle $\theta$. The frequency scale in absolute units on the right side of the plots corresponds to the unit cell dimensions used in the experiment in Section \ref{sec:meth_exp}. The  $\textnormal{MY}\Gamma\textnormal{XM}$ path corresponds to the movement along the edges of the Brillouin zone (shown in the inset). Notice the relative lack of change of frequency along the $\Gamma\textnormal{X}$ section. \textbf{b)} Normalized plasma frequency as a function of the rotation angle $\theta$. Solid line marks the simulated result obtained using COMSOL, crosses mark the points obtained experimentally.}
\label{fig:plasma}
\end{figure}

We used COMSOL Multiphysics' frequency solver to calculate the dispersive characteristics for the lowest mode of the proposed metamaterial. It is important to note that because the unit cell is elongated, the X and Y points of the Brillouin zone are not equivalent. The dispersion curves for the MY$\Gamma$XM path along the borders of the Brillouin zone are presented in Figure \ref{fig:plasma}a. The coordinates of the points within this path in the inverse space are $\Gamma(0,0)^T$, X$(\pi/b,0)^T$, Y$(0,\pi/a)^T$, M$(\pi/b,\pi/a)^T$.
  As the obround rods are rotated, the $\Gamma$ point traces a curve as shown in Figure \ref{fig:plasma}b. Based on it, we expect the maximum achievable range of plasma frequency tuning in this system to be 29.1\% ($k_0a/2\pi$ varying between 0.361 and 0.484). Here, the frequency tuning is defined as $2(f_\text{max}-f_\text{min})/(f_\text{max}+f_\text{min})$.
 


\subsection{Determination of the plasma frequency tuning using the fundamental cavity mode}

In order to experimentally demonstrate the tuning of plasma frequency we opted to use a microwave cavity filled with the obround-rod based metamaterial. By changing the plasma frequency of the metamaterial, the dielectric permittivity and the frequency of the cavity's fundamental TM110 mode are tuned as well. It is known \cite{Balafendiev2022} that the fundamental resonant frequency of the rectangular microwave cavity in the case of a square cross-section filled with a uni-axial wire medium can be calculated as:

\begin{equation}
\left(\frac{\omega_{\rm res}}{c}\right)^2 \equiv k_{\rm res}^2 
\overset{\mathrm{TM110}}{=} 
k_p^2 + 2\left(\frac{\pi}{d}\right)^2,
\label{eq:iso}
\end{equation}

where air is assumed as the host medium, $k_{\rm res}$ is the free space momentum at the resonant frequency, $k_p$ is the the plasma wave number of the metamaterial and $\pi/d=k_x=k_y$ are transverse wave vector components for the case of the TM110 resonance. However, this does not apply when the metamaterial used exhibits a degree of anisotropy in the plane transverse to the wires. In such a case the dispersion curve of the medium is characterized by an equation for an ellipsoid with coefficients $\alpha$ and $\beta$ describing the degree of anisotropy exhibited (eq \ref{eq:aniso}). These coefficients can be determined via several methods, one of which we detail in Section \ref{sec:meth_disp}. However, as can be seen in Figure \ref{fig:plasma}a, not only is the tunable medium we used clearly anisotropic, 
but for most of the range of rotation the $\Gamma$X section remains virtually flat. For this reason, approximating $\alpha=0$ and $\beta=1$ maintains a good degree of accuracy, with both methods resulting in errors less than 0.5\% when applied to the numerical results (Table \ref{tab:coeff}). As such, the relation between $k_{\rm res}$ and $k_p$ in our case can be written as:

\begin{equation}
k_{\rm res}^2 =  k_p^2 + \alpha k_x^2 + \beta k_y^2 \overset{\mathrm{\alpha=0}}{=} k_p^2 + k_y^2 = k_p^2 + \left(\frac{\pi}{d}\right)^2.
\label{eq:aniso}
\end{equation}

Using this approximation, we were able to experimentally demonstrate the tuning of the plasma frequency by deriving it from the first resonance frequency of a microwave cavity filled with the proposed metamaterial. The values found this way are plotted with black crosses in Figure \ref{fig:plasma}b. In this way, a tuning percentage of 26\% has been demonstrated.


\begin{table}[h]
    \centering
    \begin{tabular}{ccccccccc}
    \hline
    &Angle, degrees & 0 & 30 & 60 & 90 & 120 & 150 & 180 \\
    &Plasma frequency, $k_pa/2\pi$ & 0.48 & 0.47 & 0.44 & 0.40 & 0.38 & 0.37 & 0.36 \\
    \hline
    {\makecell{Ignoring anisotropy \\ $\alpha = 1, \beta = 1$}}
    &Error, \%  &  -1.16 & -1.30 & -1.67 & -2.09 & -2.38 & -2.43  & -2.36 \\
    &$\alpha$ & 0.50 & 0.33 & 0.17 & 0.07 & 0.07 & 0.13 & 0.20 \\
    &$\beta$  & 0.72 & 0.80 & 0.96 & 1.07  & 1.07  & 1.01  & 0.99 \\
    \multirow{-3}{*}{\makecell{Polynomial \\ approximation}}
    &Error, \%  &  0.00 & 0.03 & -0.10 & -0.22 & -0.25 & -0.18 & -0.17 \\
    {\makecell{No dependence on $k_x$ \\ $\alpha = 0, \beta = 1$}}
    &Error, \%  &  0.33 & 0.26 & 0.14 & 0.08 & 0.09 & 0.20  & 0.33 \\
    \hline
    \end{tabular}
    \caption{Polynomial coefficients and error values for three ways of recalculating the fundamental resonant mode frequency to plasma mode frequency (listed as the normalized plasma wave vector). Both the plasma frequency and the resonance frequency are taken from a numerical simulation.}
    \label{tab:coeff}
\end{table}

\section{Methods}

\subsection{Experimental setup} \label{sec:meth_exp}

\begin{figure}[h]
\centering
\includegraphics[width=0.6\linewidth]{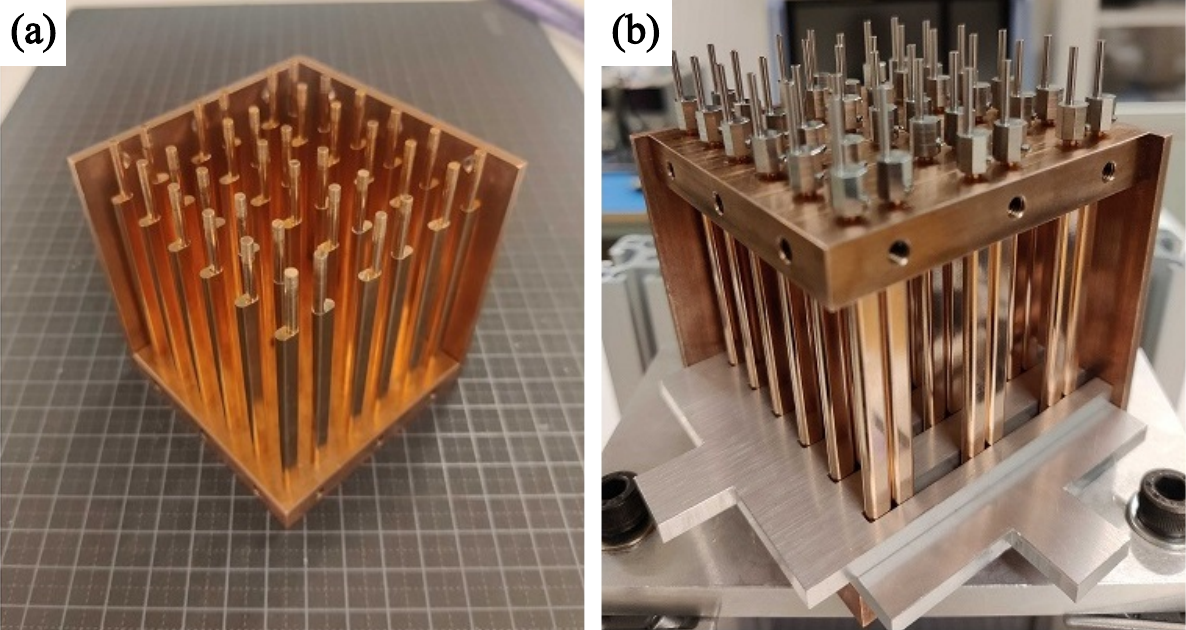}
\caption{Experimental prototype of the obround rod-based resonator. (a) Eccentric obround rods at $0^{\circ}$ rotation angle inside the cavity with the top lid and two side walls removed. (b) Same cavity with the top lid and rotation wheels placed back. The rods are rotated to $180^{\circ}$, with aluminum combs being used to further ensure the initial alignment.}
\label{fig:prototype}
\end{figure}

\begin{figure}[h]
\centering
\hspace{1.8cm}
\includegraphics[width=0.8\linewidth]{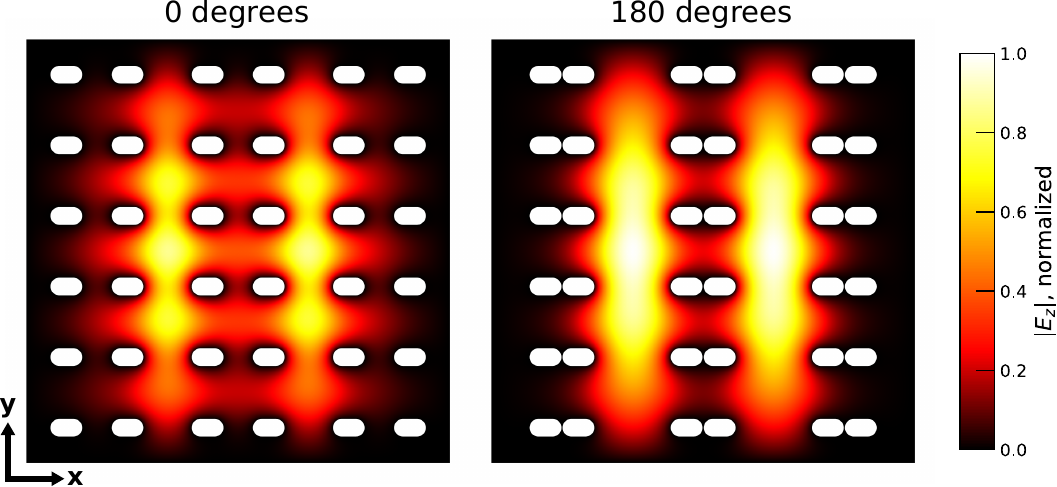}
\caption{Electric field norm distribution of the TM110 resonant mode in XY cross section. The obround rods are rotated at (a) $0^{\circ}$ and (b) $180^{\circ}$ resulting in TM110 mode at highest frequency and lowest frequency respectively in the tuning range.}
\label{fig:field_2D}
\end{figure}

We demonstrate the tuning mechanism of the proposed metamaterial using a copper cavity (see Figure \ref{fig:prototype}a). A cross-section of this cavity is shown in Figure \ref{fig:field_2D}. It consists of 18 unit cells with periods $a~=~$\SIadj{12}{\milli\meter} and $b~=~$\SIadj{24}{\milli\meter}. Each unit cell comprises 2 obround rods of length \SI{60}{\milli\meter}. The unit cells are arranged in a $3\times6$ configuration, resulting in cavity dimensions of 72~$\times$~72~$\times$~\SI{60}{\milli\meter}$^3$. The cavity can be assembled and disassembled easily, with \SIadj{10}{\milli\meter} thick top and bottom walls and \SIadj{3}{\milli\meter} thick side walls. The cavity and the rods within are made of copper with $99.90\%$ purity. The coaxial probes acting as the monopole antennas exciting the cavity were placed in the middle of the top and bottom sides so as to minimize the excitation of the TE and TEM modes. 

S-parameter measurements were performed using a Rohde\,\&\,Schwarz ZNA26A vector network analyzer (VNA), connecting the VNA to the antennas inserted into the cavity through coaxial cables with the SMA connectors. The cavity is designed in such a way that the rods can be rotated from outside without opening the resonator. The measurements in the current setup were made at a set of rotation angles using custom-made rotation wheels (at the top of the resonator, see Figure~\ref{fig:prototype}b) connected to the rods.
To compare the measurements with the ideal case, numerical simulations of the prototype model were performed using the CST Microwave Studio FEM~solver. The resulting electric field profiles of the TM110 mode for the cases of the highest and lowest mode frequencies can be seen in Figure \ref{fig:field_2D}. 

\begin{figure}[h]
\includegraphics[width=1\linewidth]{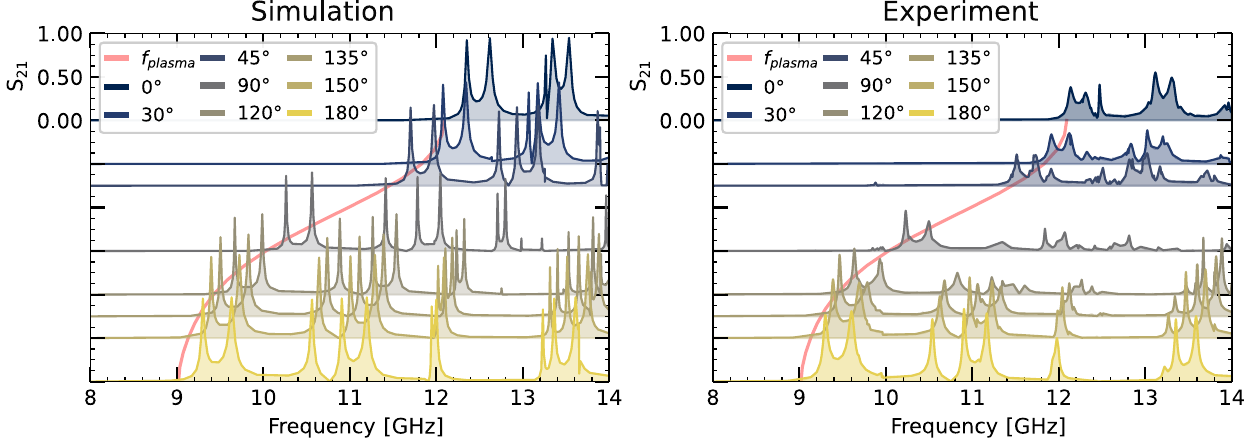}
\caption{S$_{21}$ spectra as obtained from 3D simulations in CST and VNA measurements for eight rotation angles of the rods. The spectrum corresponding to $0^{\circ}$ rotation angle is shown with exact vertical scale whereas each of the following spectra is offset by $-\theta/60$. The red curve on both plots shows the numerically calculated dependence of the plasma frequency on $\theta$.
}
\label{fig:S21}
\end{figure}

In Figure \ref{fig:S21}, the S21 transmission measurements from VNA are compared to the 3D simulation results from CST frequency domain solver at 8 rotation angles of the obround rods. As the rods are rotated from $180^{\circ}$ (lowest frequencies) to $0^{\circ}$ (highest frequencies) the entire spectrum of the cavity resonances can be seen moving lower in frequency, following the numerically calculated curve of plasma frequency tuning. 
The lowest TM resonant mode is tuned in the frequency range of 9.3 to \SI{12.14}{\giga\hertz} (tunability of $28\%$).
There is a good agreement between experimental measurements and simulations, however there are certain discrepancies which are more prominent at lower rotation angles (higher frequencies), with largest difference of 1.38\% (between \SI{12.31}{\giga\hertz} in the simulation and \SI{12.14}{\giga\hertz} in the experiment) seen at $0^{\circ}$. 
These discrepancies can be attributed to the (50-\SI{100}{\micro\metre}) leeway in the axle holes in the top and bottom plates of the cavity, added to facilitate the rotation of the rods in the prototype. The gaps between the holes and the axle create a coaxial cable-like channel leading to radiation losses.


\subsection{Dispersion contours and anisotropy} \label{sec:meth_disp}

\begin{figure*}[h]
\includegraphics[width=1\linewidth]{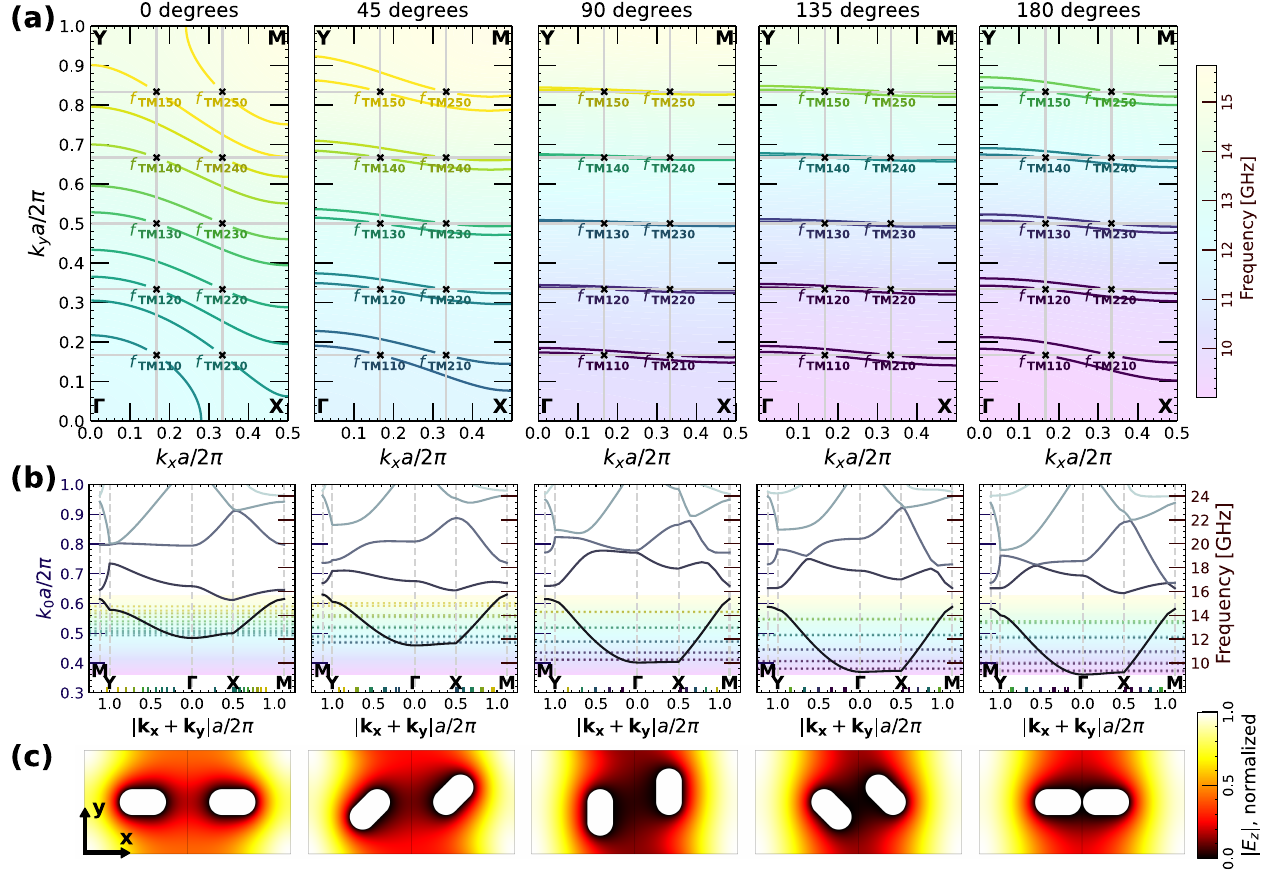}
\caption{\textbf{a)} Fundamental mode frequency surfaces within the first Brillouin zone for the cases of five rotation angles. Solid lines map out the frequencies corresponding to the quantized wave vector components of the first few TM modes of the cavity used in the experiment.\textbf{ b)} Dispersion diagrams of the first few modes for the MY$\Gamma$XM path along the edge of the Brillouin zone. Normalized absolute value of the wave vector component transverse to the rods is used for the x axis. The colored area corresponds to the frequency range of within which the first mode is tuned (also shown in a) ). The colored dotted lines show the TM mode frequencies and the colored notches along the x axis indicate where they intersect the boundary of the first Brillouin zone. \textbf{ c)} Normalized electric field profile for the longitudinal electric field component of the fundamental mode for the cases of five rotation angles.}
\label{fig:disp_full}
\end{figure*}

The straightforward derivation of the resonant frequency shown in Eq.~\ref{eq:iso} is not applicable to anisotropic material like the one presented here. In order to better characterize the degree of anisotropy exhibited by the rotating elements, we have performed a series of numerical simulations of an infinite medium using COMSOL Multiphysics' eigenmode solver. The resulting frequency contours and dispersion diagrams are shown in Figure~\ref{fig:disp_full}. 
The mode structure of the proposed cavity is not different from that of a wire metamaterial filled cavity that we have discussed in \cite{Balafendiev2022}. For a given TM$_{lmn}$ mode of the cavity of size $d_x \times d_y \times d_z$, the electric field is aligned with the rods and in a coordinate system with its origin at the center of the cavity its distribution can be written as $E_z^{lmn}=E_0\cos(k_xx)\cos(k_yy)\cos(k_zz)$, where $k_x=l\pi/d_x,k_y=m\pi/d_y$, and $k_z=n\pi/d_z$. 
The TM modes of a given cavity can be thought of as "quantizing" the frequency surfaces of an infinite medium. A grid of the ten first TM$_{lm0}$ modes can be seen superimposed with the frequency surface in Figure~\ref{fig:disp_full} with the isofrequency contours mapping out the same frequency values for different values of $k_x$ and $k_y$.

The isofrequency line for the first resonance, TM$_{110}$, already shows some noticable ellipticity at 0 degrees. This ellipticity is inherent to the modified unit cell with off-center elements that we are using \cite{sakhno2024}. As the rods rotate, it flattens out even further, becoming virtually indistinguishable from TM$_{210}$ at 90 degrees. Evidently, at those angles the frequency is nearly independent of $k_x$. The middle row shows the same change of the first resonance frequency of the infinite medium along the edges of the first Brillouin zone, as well as the frequency dependence of the higher-order modes. The range of frequencies covered by the top row subfigures is visualised with the colored gradient, with the frequencies of the cavity resonances shown with the dotted lines. The colored notches along the vertical axis note the points where isofrequency contours intersect the edge of the first Brillouin zone. The same independence of the resonance frequency from the x-component of the wavevector can be observed through the flattening of the dispersion curve at the $\Gamma$X section of the plot. The bottom row shows the positions of the obround rods and the electric field distributions corresponding to each angle value shown.

In order to characterize the degree of anisotropy exhibited by the proposed metamaterial we have used 4th degree polynomial approximation of the two dispersion surface cuts: along $k_y=0$ ($\Gamma$X path) and along $k_x=0$ ($\Gamma$Y path). The values found (listed in Table \ref{tab:coeff}) have confirmed our observations that near the $\Gamma$ point the dependence on $k_x$ can be largely ignored, while the dependence on $k_y$ stays virtually the same. While this does not hold as strongly for the higher order TM modes that are further removed from the $\Gamma$ point, for the TM110 mode it does give us an easy way to convert the resonance frequency into the plasma frequency.

\section{Summary}


In this paper we have investigated several geometries of a wire-metamaterial inspired medium with a plasma frequency tunable via rotational motion. We have found that in metamaterial with a single rotating element per unit cell, a tuning range of 11.7\% is theoretically possible. Extending the unit cell to include two rotating elements has been shown to significantly extend this tuning range, with a tuning range of 26\% being demonstrated experimentally and a theoretical maximum range of 29\% shown via numerical simulations.
Modes of a microwave cavity were used to indirectly show the change in the plasma frequency. The dispersive properties of the proposed metamaterial were investigated to facilitate this. In the process, the proposed metamaterial has been shown to have a degree of anisotropy in the transverse plane, with the resonance frequency being virtually independent of the wavevector component along the long side of the unit cell. Subsequently, the plasma frequency was obtained from the TM110 mode frequency of the resonator.

The proposed metamaterial presents a way to dynamically tune the frequency of the ENZ condition within a microwave range, something that can not be achieved with natural materials. This tunable ENZ can be useful for a number of microwave applications \cite{silveirinha_tunneling_2006, enoch_metamaterial_2002, alu_epsilon-near-zero_2007, zhou_broadband_2018, li_geometry-independent_2022}. Additionally, the discussed case of a metamaterial changing the resonant frequency of a cavity is important in the rapidly developing direction of dark matter search using plasma haloscopes \cite{Lawson19, Millar23, Balafendiev2022, Sikivie83}. In this context, the proposed metamaterial combines the benefits of using a WM to enable high-volume GHz-range haloscopes with an established mechanical solution of using a rotating element for frequency tuning \cite{Boutan18, Zhong18}. A more detailed analysis of the applicability of such haloscopes will be carried out in future work.

\medskip
\section*{Acknowledgments}
This research is supported by the Swedish Research Council(VR) under Dnr 2019-02337 "Detecting Axion Dark Matter In The Sky And In The Lab" (AxionDM). The authors gratefully acknowledge the funding support from the Knut and Alice Wallenberg Foundation. Fermilab is operated by Fermi Forward Discovery Group, LLC under Contract No. 89243024CSC000002 with the U.S. Department of Energy, Office of Science, Office of High Energy Physics. We thank members of the ALPHA Collaboration for many helpful discussions and Junu Jeong in particular for his valuable comments on the manuscript.

\section*{Conflict of Interest}
The authors declare no conflict of interest.
\medskip

%


\bibliographystyle{MSP}
\bibliography{bibliography}






\end{document}